\documentclass{ws-procs9x6}

\begin{document}
\newcommand{\hami}{\mathcal{H}}
\newcommand{\nhat}{\hat{n}}
\newcommand{\Zhat}{\hat{Z}}
\newcommand{\Phat}{\hat{P}}
\newcommand{\ahat}{\hat{a}}
\newcommand{\ahatdag}{\hat{a}^\dagger}
\newcommand{\neff}{N_{\mbox{\scriptsize{eff}}}}
\newcommand{\nmax}{n_{\mbox{\scriptsize{max}}}}
\newcommand{\dpc}{\Delta_{pc}}
\newcommand{\Zho}{Z_{\mbox{\scriptsize{ho}}}}
\newcommand{\Fhatopto}{\hat{\mathcal{F}}_{\mbox{\scriptsize{opto}}}}
\newcommand{\dca}{\Delta_{ca}}

\newcommand{\snn}{S_{nn}}
\newcommand{\snnplus}{S_{nn}^{(+)}}
\newcommand{\snnminus}{S_{nn}^{(-)}}
\newcommand{\snnplusminus}{S_{nn}^{(\pm)}}

\title{Quantum micro-mechanics with ultracold atoms}

\author{THIERRY BOTTER$^1$, DANIEL BROOKS$^1$, SUBHADEEP GUPTA$^2$, ZHAO-YUAN MA$^1$, KEVIN L. MOORE$^1$, KATER W. MURCH$^1$, TOM P. PURDY$^1$ and DAN M. STAMPER-KURN$^{1,3,*}$}

\address{$^1$ Department of Physics, University of California, Berkeley, CA  94720, USA \\
$^2$ Department of Physics, University of Washington, Seattle, WA  98195, USA \\
$^3$ Materials Sciences Division, Lawrence Berkeley National Laboratory, Berkeley, CA  94720, USA \\
$^*$E-mail: dmsk@berkeley.edu \\
physics.berkeley.edu/research/ultracold}

\begin{abstract}
In many experiments isolated atoms and ions have been inserted into
high-finesse optical resonators for the study of fundamental quantum
optics and quantum information.  Here, we introduce another
application of such a system, as the realization of cavity
optomechanics where the collective motion of an atomic ensemble
serves the role of a moveable optical element in an optical
resonator.  Compared with other optomechanical systems, such as
those incorporating nanofabricated cantilevers or the large cavity
mirrors of gravitational observatories, our cold-atom realization
offers direct access to the quantum regime. We describe experimental
investigations of optomechanical effects, such as the bistability of
collective atomic motion and the first quantification of measurement
backaction for a macroscopic object, and discuss future directions
for this nascent field.
\end{abstract}


\bodymatter

\section{Introduction}\label{sec:intro}

Cavity opto-mechanics describes a paradigmatic system for quantum
metrology: a massive object with mechanical degrees of freedom is
coupled to and measured by a bosonic field. Interest in this generic
system is motivated by several considerations.  For one, the system
allows one to explore and address basic questions about quantum
limits to measurement. In this context, quantum limits to quadrature
specific and non-specific measurements, both for those performed
directly on the mechanical object and also those performed through
the mediation of an amplifier \cite{brag95qmbook}. Second, as a
detectors of weak forces, cavity opto-mechanical systems in the
quantum regime may yield improvements in applications ranging from
the nanoscale (e.g.\ for atomic or magnetic force microscopies) to
the macroscale (e.g.\ in ground- or space-based gravity wave
observatories).  Finally, such systems, constructed with ever-larger
mechanical objects, may allow one to test the validity of quantum
mechanics for massive macroscopic objects.  Striking developments in
this field were presented at ICAP 2008 by Harris and Kippenberg.

Our contribution to this developing field is the realization that a
cavity opto-mechanical system can be constructed using a large gas
of ultracold atoms as the mechanical object.  Having developed an
apparatus that allows quantum gases to be trapped within the optical
mode of a high-finesse Fabry-Perot optical resonator, we are now
able to investigate basic properties of opto-mechanical systems.
Several of these investigations are described below.  The
atoms-based mechanical oscillator may be considered small by some,
with a mass ($\simeq 10^{-17}$ g) lying geometrically halfway
between the single-atom limit explored at the quantum regime in ion
and atom traps ($10^{-22}$ g) \cite[for example]{meek96,bouc99}, and
the small ($\simeq 10^{-12}$ g) nanofabricated systems now
approaching quantum limits \cite{laha04approach,rega08}.
Nevertheless, our system offers the advantages of immediate access
to the quantum mechanical regime, of the \emph{ab initio}
theoretical basis derived directly from quantum optics and atomic
physics, and of the tunability and amenability to broad new probing
methods that are standard in ultracold atomic physics.  Our
motivation for probing cavity opto-mechanics with our setup is not
just to poach the outstanding milestones of this field (e.g.\
reaching the motional ground state or observing measurement
backaction and quantum fluctuations of radiation pressure with a
macroscopic object \cite{murc08backaction}).  Rather, we hope to
contribute to the development of macroscopic quantum devices by
clarifying experimental requirements and the role of and limits to
technical noise, developing optimal approaches to signal analysis
and system control, exploring the operation and uses of multi-mode
quantum devices, and defining different physical regimes for such
systems. Also, our opto-mechanical system may have direct
application as part of an atom-based precision (perhaps
interferometric) sensor.

\section{Collective modes of an intracavity atomic ensemble}

The theoretical reasoning for considering a trapped atomic gas
within a high-finesse optical resonator as a macroscopic cavity
opto-mechanical system is laid out in recent work
\cite{murc08backaction}. Recapping that discussion, we consider the
dispersive coupling of an ensemble of $N$ identical two-level atoms
to a single standing-wave mode of a Fabry-Perot cavity, obtaining
the spectrum of ``bright'' eigenstates of the atoms-cavity system
according to the following Hamiltonian:
\begin{equation}
\hami = \hbar \omega_c \nhat + \sum_{i} \frac{\hbar g^2(z_i)}{\dca}
+ \hami_{a} + \hami_{in/out}
\end{equation}
Here $\nhat$ is the cavity photon number operator, $\dca = \omega_c
- \omega_a$ is the difference between the empty-cavity and atomic
resonance frequencies, and $g(z_i) = g_0 \sin(k_p z)$ is the
spatially dependent atom-cavity coupling frequency with $z_i$ being
the position of atom $i$ and $k_p$ being the wavevector at the
cavity resonance.  The term $\hami_a$ describes the energetics of
atomic motion while $\hami_{in/out}$ describes the electromagnetic
modes outside the cavity.  Note that this expression already treats
the atom-cavity coupling to second order in $g$. Repeating this
analysis starting from the first-order term does not change our
conclusions substantially.

Now, let us assume that all the atoms are trapped in harmonic
potentials with ``mechanical'' trap frequency $\omega_z$ and neglect
motion along directions other than the cavity axis.  Further, we
treat the atomic motion only to first order in atomic displacements,
$\delta z_i$, from their equilibrium positions, $\bar{z}_i$;  i.e.\
we assume atoms to be confined in the Lamb-Dicke regime with $k_p
\delta z_i \ll 1$. We now obtain the canonical cavity
opto-mechanical Hamitonian \cite{kipp08sciencereview} as
\begin{equation}
\hami = \hbar \omega_c^\prime \nhat + \hbar \omega_z \ahatdag \ahat
- F \Zhat \nhat + \hami^\prime_{a} + \hami_{in/out} \label{eq:hami}
\end{equation}

We make several steps to arrive at this expression. First, we allow
the cavity resonance frequency to be modified as $\omega_c^\prime =
\omega_c + \sum_i g^2(\bar{z}_i) / \dca$, accounting for the cavity
resonance shift due to the atoms at their equilibrium positions.
Second, we introduce the collective position variable $\Zhat =
\neff^{-1} \sum_i \sin(2 k_p \bar{z}_i) \delta z_i$ that, along with
a weighted sum $\Phat = \sum_i \sin(2 k_p \bar{z}_i) p_i$ of the
atomic momenta $p_i$, describes the one collective motion within the
atomic ensemble that is coupled to the cavity-optical field. The
operators $\ahat$ and $\ahatdag$ are defined conventionally for this
mode. In our treatment, absent the presence of light within the
optical cavity, this mode is harmonic, oscillating at the mechanical
frequency $\omega_z$, and endowed with a mass $M$ equal to that of
$\neff = \sum_i \sin^2(2 k_p \bar{z}_i)$ atoms. Third, we summarize
the opto-mechanical coupling by the per-photon force $F = \neff
\hbar k g_0^2 / \dca$ that acts on the collective mechanical mode.
Finally, we lump all the remaining atomic degrees of freedom, and
also the neglected higher order atom-cavity couplings, into the term
$\hami^\prime_a$.

With this expression in hand, we may turn immediately to the
literature on cavity opto-mechanical systems to identify the
phenomenology expected for our atoms-cavity system.  Several such
phenomena are best described by referring to the opto-mechanical
force on the collective atomic mode, given as
\begin{equation}
\Fhatopto = - M \omega_z^2 \Zhat + F \nhat. \label{eq:optoforce}
\end{equation}

We consider the following effects:
\begin{itemize}
\item{} If we allow the state of the cavity to follow the atomic motion adiabatically ($\omega_z \ll \kappa$),
neglect quantum-optical fluctuations of the cavity field, and assume
the collective atomic displacement remains small, the linear
variation of $\langle \nhat \rangle$ with $\Zhat$ modifies the
vibration frequency of the collective atomic motion.  Here, $\kappa$
is the cavity half-linewidth.  This modification, known as the
``optical spring,'' has been observed in various opto-mechanical
systems and has been used to trap macroscopic objects optically
\cite{shea04spring,corb06spring,corb07gram}. We have made
preliminary observations of the optical spring effect in our system
as well.

\item{} For larger atomic displacements, the opto-mechanical force
may become notably anharmonic, and even, under suitable conditions,
bistable \cite{dors83bistability}.  Our observations of the
resulting opto-mechanical bistability \cite{gupt07nonlinear} are
discussed in Sec.\ \ref{sec:bistab}.

\item{} When the cavity field no longer follows the atomic motion
adiabatically, the opto-mechanical potential is no longer
conservative.  The dramatic effects of such non-adiabaticity are the
cavity-induced damping or coherent amplification of the mechanical
motion \cite{brag67}.  Such effects of dynamical backaction have
been detected in several micro-mechanical systems
\cite{kipp05anal,arci06,giga06cooling} and also for single
\cite{maun04cooling} or multiple atoms \cite{chan03} trapped within
a cavity.

\item{} Finally, we consider also the effects of quantum-optical
fluctuations of the intracavity photon number and, thereby, of the
optical forces on the atomic ensemble.  It can be shown that these
force fluctuations represent the backaction of quantum measurements
of the collective atomic position \cite{murc08backaction}, as
described in Sec.\ \ref{sec:backaction}.
\end{itemize}

\section{Collective atomic modes in various regimes}

\begin{figure}
\begin{center}
\psfig{file=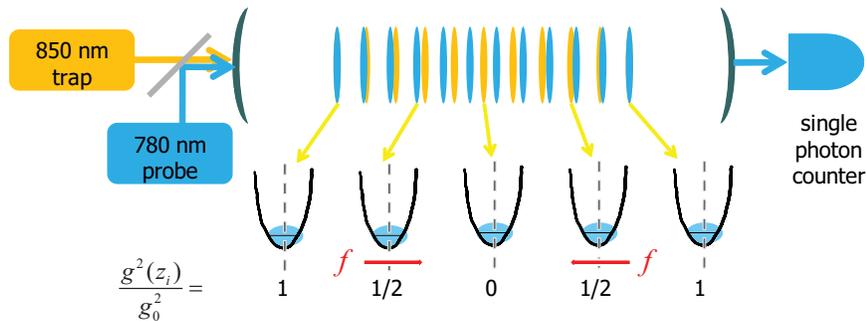,width=4.5in}
\end{center}
\caption{Scheme for opto-mechanics with ultracold atoms in the
Lamb-Dicke confinement regime.  A high finesse cavity supports two
longitudinal modes -- one with wavelength of about 780 nm that is
near the D2 resonance of $^{87}$Rb atoms trapped within the
resonator, and another with wavelength of about 850 nm.  Light at
the 850 nm resonance produces a one-dimensional optical lattice,
with trap minima indicated in orange, in which atoms are confined
within the lowest vibrational band.  These atoms induce frequency
shifts on the 780 nm cavity resonance.  The strength of this shift,
and of its dependence on the atomic position, varies between the
different sites of the trapping optical lattice, as shown.
Nevertheless, in the Lamb-Dicke confinement regime, the complex
atoms-cavity interactions reduce to a simple opto-mechanical
Hamiltonian wherein a single collective mode of harmonic motion,
characterized by position and momentum operators $\Zhat$ and
$\Phat$, respectively, is measured, actuated, and subjected to
backaction by the cavity probe.} \label{fig:scheme}
\end{figure}

The theoretical treatment described above is suitable in the
Lamb-Dicke regime of atomic confinement and under the condition that
the linear opto-mechanical coupling term ($F \Zhat \nhat$) is
dominant (i.e.\ that the intracavity atomic gas is not tuned to
positions of exclusively quadratic sensitivity).  These conditions
are met in our experiments at Berkeley, where an ultracold gas of
about $10^5$ atoms of $^{87}$Rb is transported into the mode volume
of a high-finesse Fabry-Perot optical resonator. The resonator
length is tuned so that the resonator supports one TEM$_{00}$ mode
with wavelength $\lambda_T = 850$ nm (trapping light) and another
within a given detuning $\dca$ (in the range of 100's of GHz) of the
D2 atomic resonance line (probe light). Laser light with wavelength
$\lambda_T$ is sent through the cavity to generate a 1D optical
lattice potential in which the cold atomic gas is trapped (Fig.\
\ref{fig:scheme}). The gas is strewn across over $> 100$ contiguous
sites in this 1D optical lattice. Within each well, atoms are
brought by evaporative cooling to a temperature $T \sim 700$ nK. At
this temperature, the atoms lie predominantly in the ground state of
motion along the cavity axis, with $\hbar \omega_z / k_B \simeq 2 \,
\mu\mbox{K} \gg T$, and the Lamb-Dicke condition is satisfied with
respect to the wavevector of probe light ($k_p \simeq 2 \pi / (780
\, \mbox{nm})$) used to interrogate the atomic motion.

The opto-mechanics picture of atomic motion in cavity QED has also
been considered recently by the Esslinger group in Z\"{u}rich
\cite{bren08optonote}. There, a continuous Bose-Einstein condensate
of $^{87}$Rb is trapped in a large-volume optical trap within the
cavity volume. Yet, in spite of the stark differences in the
external confinement and the motional response of the condensed gas,
a similar opto-mechanical Hamiltonian emerges.  In our prior
description of the Lamb-Dicke regime, optical forces due to cavity
probe light are found to excite and, conversely, to make the cavity
sensitive to a specific collective motion in the gas.  In the case
of a continuous condensate, the cavity optical forces excite atoms
into a specific superposition of the $\pm 2 \hbar k_p$ momentum
modes. Interference between these momentum-excited atoms and the
underlying condensate creates a spatially (according to $k_p$) and
temporally (according to the excitation energy) periodic density
grating that is sensed via the cavity resonance frequency.  Thus,
identifying operators $\ahat$ and $\ahatdag$ with this
momentum-space excitation and the operator $\Zhat$ with the density
modulation, we arrive again at the Hamiltonian of Eq.\
\ref{eq:hami}.

\begin{figure}
\begin{center}
\psfig{file=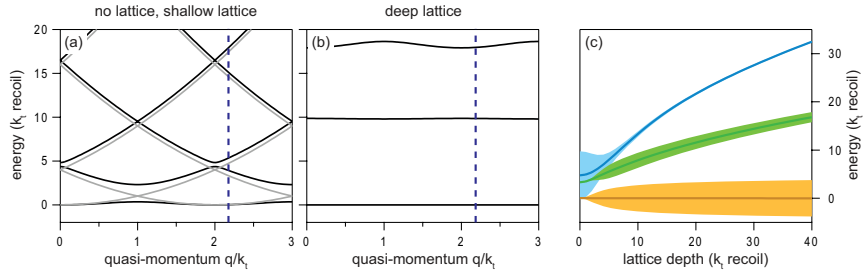,width=4.5in}
\end{center}
\caption{Influence of band structure on the opto-mechanical response
of an ultracold atomic gas confined within a high-finesse
Fabry-Perot optical resonator.  We consider the relevant macroscopic
excitation produced by cavity probe light at wavevector $k_p = 2 \pi
/ (780 \, \mbox{nm})$ within a Bose gas confined within a
one-dimensional optical lattice formed by light at wavevector $k_t =
2 \pi / (850 \, \mbox{nm})$ and with variable depth.  The gas is
cooled to zero temperature, non-interacting, and extended evenly
across many lattice sites.  Energies are scaled by the recoil energy
$E_r = \hbar^2 k_t^2 / 2 m$ and wavevectors by $k_t$.  (a) With a
weak lattice applied ($2 E_r$), the band structure for atomic
excitations (black lines) is slightly perturbed from the
free-particle excitations in the absence of a lattice (gray).  The
cavity probe excites atoms primarily to states with quasi-momenta
$\pm 2 k_p$ within the second excited band, corresponding closely to
momentum eigenstates in the lattice-free regime.  (b) In a deep
lattice ($15 E_r$), energy bands show little dispersion and are
spaced by energies scaling as the square root of the lattice depth.
(c) Lines show the energies of the three lowest energy states at
quasi-momentum $2 k_p$ as a function of the lattice depth.  The
relative probability for excitation by cavity probe light to each of
these states, taken as the square of the appropriate matrix element,
is shown by the width of the shaded regions around each line.  At
zero lattice depth, cavity probe light excites the second excited
band exclusively.  At large lattice depth, the excitation
probability to the first excited band grows while excitation to
higher bands is suppressed. At intermediate lattice depths, several
excited states are populated, indicating the onset of complex
multi-mode behaviour.} \label{fig:bands}
\end{figure}

We can attempt to bridge these two opto-mechanical treatments by
tracking the response of an extended atomic gas to spatially
periodic optical forces (due to probe light at wavelength 780 nm) as
we gradually turn up the additional optical lattice potential (due
to trapping light at wavelength 850 nm).  In the absence of the
lattice potential, a zero-temperature Bose gas forms a uniform
Bose-Einstein condensate.  The excitations of this system are
characterized by their momentum and possess an energy determined by
the Bogoliubov excitation spectrum; in Fig.\ \ref{fig:bands}(a), we
present this spectrum as a free-particle dispersion relation,
neglecting the effects of weak interatomic interactions. The
spatially periodic optical force of the cavity probe excites a
superposition of momentum excitations as described above.

Adding the lattice potential changes both the state of the
Bose-Einstein condensate, which now occupies the lowest Bloch state,
and also the state of excitations, which are now characterized by
their quasi-momentum and by the band index.  There are now many
excitations of the fluid that may be excited at the quasi-momentum
selected by the spatially periodic cavity probe.  In the case that
the lattice is very shallow, shown in Fig.\ \ref{fig:bands}(a), the
cavity probe will still populate only one excited state nearly
exclusively.  Given the relation between the wavelengths of the
trapping (850 nm) and cavity-probe light (780 nm), this excited
state lies in the second excited band.  As the lattice is deepened,
however, matrix elements connecting to quasi-momentum states on
other bands will grow (shown in Fig.\ \ref{fig:bands}(c)).  Now our
simple opto-mechanical picture is made substantially more complex,
with multiple mechanical modes oscillating with differing mechanical
frequencies all influencing the optical properties of the cavity.

Continuing to deepen the optical lattice this complexity will be
alleviated when we reach the Lamb-Dicke regime, i.e.\ as the
Lamb-Dicke parameter $k_p \delta z$ becomes ever smaller, the
probabilities of excitation from the ground state via the cavity
probe hone in on the first excited band.  We calculate such
probabilities as $p_i \propto \left| <2 k_p; i| \cos(2 k_p z) | g>
\right|^2$ where the ket is the $2 k_p$ quasi-momentum Bloch state
in the $i$th band, and the bra is the ground state in the lattice
considered. Here, we interpret excitations to higher bands as being
controlled by terms of higher order in the Lamb-Dicke parameter,
e.g.\ excitations to the second excited band result from couplings
that are quadratic in the atomic positions.

Thus, we confirm that a simple opto-mechanics picture emerges for
the collective atomic motion within a cavity both in the shallow-
and deep-lattice limits.  We note, however, that these limits differ
in two important ways.  First, we see that the mechanical
oscillation frequency for the collective atomic motion is
constrained to lie near the bulk Bragg excitation frequency in the
shallow-lattice limit, whereas it may be tuned to arbitrarily high
frequencies (scaling as the square root of the lattice depth) in the
deep-lattice limit.  The ready tunability of the mechanical
frequency in the latter limit may allow for explorations of quantum
opto-mechanical systems in various regimes, e.g.\ in the the
resolved side-band regime where ground-state cavity cooling and also
quantum-limited motional amplification are possible
\cite{marq07sideband,wils07groundstate}. Second, we see that the
mechanical excitation frequency has a significant quasi-momentum
(Doppler) dependence in the shallow-lattice limit. This dependence
makes it advantageous to use low-temperature Bose-Einstein
condensates for experiments of opto-mechanics, as indeed achieved in
the Z\"{u}rich experiments, so as to minimize the Doppler width of
the Bragg excitation frequency. In contrast, the excitation
bandwidth is dramatically reduced (exponentially with the lattice
depth) in the deep-lattice limit. Thus, one can conduct
opto-mechanics experiments with long-lived mechanical resonances in
the deep-lattice limit without bothering to condense the atomic gas.
Nevertheless, we note that variations in the mechanical frequency
due to the presence of significant radial motion (not considered in
this one-dimensional treatment) do indeed limit the mechanical
quality factor in the Berkeley experiments.

\section{Effects of the conservative optomechanical potential: optomechanical bistability}
\label{sec:bistab}

The observation of cavity nonlinearity and bistability arising from
collective atomic motion is described in recent work
\cite{gupt07nonlinear}.  Briefly, we find that the optical force due
to cavity probe light will displace the equilibrium collective
atomic position $\langle \Zhat \rangle$, leading to a
probe-intensity-dependent shift of the cavity resonance frequency.
By recording the cavity transmission as cavity probe light was swept
across the cavity resonance, we observed asymmetric and shifted
cavity resonance lines, and also hallmarks of optical bistability.

Refractive optical bistability is well studied in a variety of
experimental systems \cite{boyd03}.  One unique aspect of our
experiment is the observation of both branches of optical
bistability at average cavity photon numbers as low as $0.02$.  The
root of such strong optical nonlinearities is the presence within
the cavity of a medium that (1) responds significantly to the
presence of infrequent cavity photons (owing to the strong
collective cooperativity) and (2) recalls the presence of such
photons for long coherence times (here, the coherence is stored
within the long-lived collective motion of the gas).  It is
interesting to consider utilizing such long-lived motional
coherence, rather than the shorter-lived internal state coherence
typically considered, for the various applications of cavity QED and
nonlinear optics in quantum information science, e.g.\ photon
storage and generation, single-photon detection, quantum logic
gates, etc.

Such motion-induced cavity bistability can also be understood in the
context of the opto-mechanical forces described by Eq.\
\ref{eq:optoforce}.  Neglecting the non-adiabatic following of the
cavity field to the collective motion (essentially taking $\kappa /
\omega_z \rightarrow \infty$ so that dynamical backaction effects
are neglected) and also the quantum fluctuations of the cavity
field, we may regard atomic motion in an optically driven cavity to
be governed by an opto-mechanical potential of the form
\begin{equation}
U(Z) = \frac{1}{2} M \omega_z^2 Z^2 + \nmax \hbar \kappa \,
\arctan\left( \frac{\dpc - F Z / \hbar}{\kappa}\right)
\label{eq:ucoll}
\end{equation}
Here $\dpc$ is the detuning of the constant frequency probe from the
modified cavity resonance frequency $\omega_c^\prime$, and $\nmax$
is the average number of cavity photons when the cavity is driven on
resonance.

\begin{figure}
\begin{center}
\psfig{file = 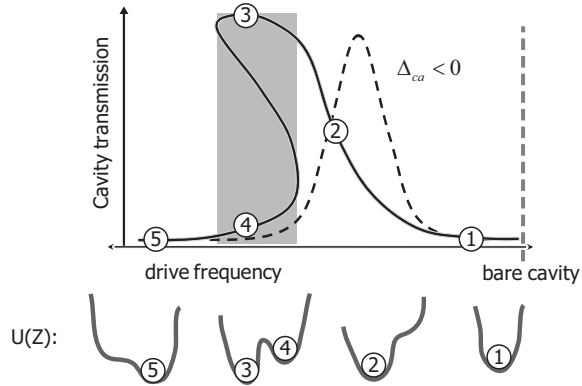, height =2in}
\end{center}
\caption{\small For different stable regimes of cavity operation,
the cavity-relevant collective mode of the intracavity atomic
ensemble is trapped in a particular minimum of the effective
potential $U(Z)$ (bottom).  In the regime of bistability, the two
stable cavity states reflect the presence of two potential minima.}
\label{fig:bistablepotential}
\end{figure}

The form of this potential is sketched in Fig.\
\ref{fig:bistablepotential} for different operating conditions of
the atoms-cavity system.  Cavity bistability \cite{gupt07nonlinear}
is now understood as reflecting an effective potential for the
collective atomic variable $Z$ that has two potential minima.
Remarkably, these potential minima may be separated by just
nanometer-scale displacements in $Z$.  Even though the inherent
quantum position uncertainty of each individual atom (10's of nm) is
much larger than this separation, the reduced uncertainty in the
collective variable $Z$ allows for these small displacements to
yield robust and distinct experimental signatures in the cavity
transmission.

\section{Quantum fluctuations of the optomechanical potential: measurement backaction}
\label{sec:backaction}

Aside from the conservative forces described above, the intracavity
atomic medium is subject also to dipole force fluctuations arising
from the quantum nature of the intracavity optical field. Indeed,
should these force fluctuations be especially large, the picture of
cavity optical non-linearity and bistability described in the
previous section, in which we implicitly assume that the collective
atomic motion may follow adiabatically into a local minimum of an
opto-mechanical potential, must be dramatically modified.  To assess
the strength of such force fluctuations, let us consider the impact
on the atomic ensemble of a single photon traversing the optical
cavity.  During its residence time of $\sim 1 / 2 \kappa$, such a
photon would cause a dipole force that imparts an impulse of $\Delta
P = f/(2 \kappa)$ on the atomic medium, following which the
collective mode is displaced by a distance $\Delta Z = \Delta P / (M
\omega_z)$; in turn, this displacement will shift the cavity
resonance frequency by $F \Delta Z / \hbar$.  Comparing this
single-photon-induced, transient frequency shift with the cavity
half-linewidth leads us to define a dimensionless ``granularity
parameter'' as
\begin{equation}
\epsilon = \sqrt{ \frac{F \Delta Z}{\hbar \kappa}} = \frac{F
\Zho}{\hbar \kappa},
\end{equation}
where $\Zho = \sqrt{\hbar / 2 M \omega_z}$ is the harmonic
oscillator length for the atomic collective mode.  The condition
$\epsilon > 1$ marks the granular (or strong) opto-mechanical
coupling regime in which the disturbance of the collective atomic
mode by single photons is discernible both in direct quantum-limited
measurements of the collective atomic motion and also in subsequent
single-photon measurements of the cavity resonance frequency.  In
our experiments, the granularity parameter is readily tuned by
adjusting frequency difference between the cavity and atomic
resonance $\Delta_{ca}$.  Under conditions of our recent work, the
granular regime is reached at  $|\dca|/(2\pi) < 27$ GHz.

In recent work, we have focused on effects of fluctuations of the
dipole force in the non-granular regime, attained at atom-cavity
detunings in the 100 GHz range.  As described in our work
\cite{murc08backaction}, and also derived in earlier treatments
\cite{hora97,vule00,murr06,marq07sideband}, these fluctuations will
cause the motional energy of the collective atomic mode to vary
according to the following relation:
\begin{equation}
\frac{d}{d t} \langle a^\dagger a \rangle = \kappa^{2} \epsilon^2
\left[\snnminus + \left(\snnminus - \snnplus\right) \langle
a^\dagger a \rangle \right] \label{eq:adaga}
\end{equation}
Here, the relevant dipole force fluctuations are derived from the
spectral density of intracavity photon number fluctuations at the
mechanical frequency $\omega_z$, calculated for a
coherent-state-driven cavity as $\snnplusminus = 2 \langle n \rangle
\kappa (\kappa^{2} + (\dpc \pm \omega_z)^{2})^{-1}$.  Eq.\
\ref{eq:adaga}, which can be derived readily from a rate-equation
approach \cite{marq07sideband}, reveals two manners in which the
mechanical oscillator responds to a cavity optical probe: momentum
diffusion, which raises the mechanical oscillator energy at a
constant rate, and the dynamic backaction effects of cavity-based
cooling or amplification of the mechanical motion, described by an
exponential gain or damping.

The mechanical momentum diffusion in an opto-mechanical system plays
the essential metrological role of providing the backaction
necessary in a quantum measurement, as discussed, for example, by
Caves in the context of optical interferometry \cite{cave81}.  For
$\omega_z \ll \kappa$, we see that, at constant circulating power in
the cavity, this diffusion is strongest for probe light at the
cavity resonance and weaker away from resonance.  This dependence on
the intensity and detuning of the cavity probe light precisely
matches the rate of information carried by a cavity optical probe on
the state of the mechanical oscillator.  To elucidate this point, we
recall that, under constant drive by a monochromatic input field,
the intracavity electric field oscillates at the input field
frequency with complex amplitude $E_{cav} = \eta / (\kappa - i
\dpc)$.   A displacement by $\Delta Z$ of the mechanical oscillator
varies the probe-cavity detuning by $F \Delta Z / \hbar$. In
response, the electric field in the cavity varies as
\begin{equation}
E_{cav} \simeq E_0 \left( 1 + \frac{i}{\kappa - i \dpc} \frac{F
\Delta Z}{\hbar} \right) = E_0 + E_{sig},
\end{equation}
where $E_0$ is the cavity field with the cantilever at its
equilibrium position and we expand to first order in $\Delta Z$. The
sensitivity of the cavity field to the cantilever displacement, at
constant intracavity intensity (constant $E_0$), is determined by
the magnitude of $|E_{sig}/E_0|^2 \propto 1 / (1 + \dpc^2 /
\kappa^2)$; this functional dependence matches that of the momentum
diffusion term, supporting its representing  measurement backaction.

To measure this backaction heating, we take advantage of several
features of our experiment.  First, by dint of the low temperature
of our atomic ensemble, we ensure that the effects of dynamical
backaction (cooling and amplification) are negligible.  Second, the
low quality-factor of our mechanical oscillator ensures that the
momentum diffusion of the collective atomic motion leads to an
overall heating of the atomic ensemble, allowing us to measure this
diffusion bolometrically.  Third, the large single-atom
cooperativity in our cavity QED system implies that this backaction
heating of the entire atomic ensemble dominates the single-atom
heating due to atomic spontaneous emission.  And, fourth, owing to
the finite, measured depth of our intracavity optical trap,
backaction heating can be measured via the light-induced loss rate
of atoms from the trap.  The measured light-induced heating rate was
found to be in good agreement with our predictions, providing the
first quantification of measurement backaction on a macroscopic
object at a level consistent with quantum metrology limits.

\section{Future developments: cavity QED/atom chips}

While continuing explorations of quantum opto-mechanics in our
existing apparatus, we are also developing an experimental platform
that integrates the capabilities of single- and many-atom cavity QED
onto microfabricated atom chips.  Similar platforms have been
developed recently by other groups \cite{tepe06res,colo07}. Aside
from enabling myriad applications in quantum atom optics and atom
interferometry, we anticipate the cavity QED/atom chip to provide
new capabilities in cold-atoms-based opto-mechanics. For instance,
the tight confinement provided by microfabricated magnetic traps
will allow atomic ensembles to be confined into single sites of the
intracavity optical lattice potential, providing a means of tuning
the opto-mechanical coupling between terms linear or quadratic in
$\Zhat$.  As emphasized by Harris and colleagues
\cite{thom08membrane}, a purely quadratic coupling may allow for
quantum non-demolition measurements of the energy of the macroscopic
mechanical oscillator.

\begin{figure}
\begin{center}
\psfig{file = 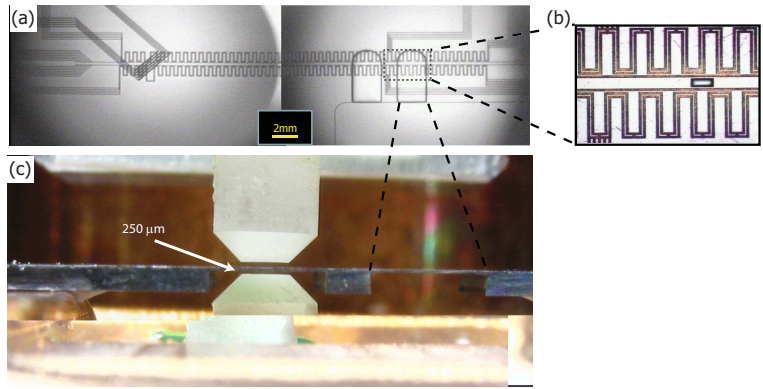, height = 2 in}
\end{center}
\caption{An integrated cavity QED/atom chip. (a) A top view of the
microfabricated silicon chip shows etched trenches, later
electroplated with copper and used to tailor the magnetic field
above the chip surface.   The left portion of the image shows wire
patterns used for producing the spherical-quadrupole field of a
magneto-optical trap and also the Ioffe-Pritchard fields for
producing stable magnetic traps.  Serpentine wires spanning the
entire chip form a magnetic conveyor system to translate atoms to
the optical cavities that are located in the right half of the
image.  (b) A detailed view shows the serpentine wires and also a
two-wire waveguide surrounding a central, rectangular hole that
pierces the atom chip.  (c) Fabry-Perot cavities are formed by
mirrors straddling the atom chip. Between the mirrors, the chip is
thinned to below 100 $\mu$m -- outlines of the thinned areas are
seen also in (a).  The cavity mode light passes unhindered through
the chip via the microfabricated holes shown in (b).}
\label{fig:atomchip}
\end{figure}

\bibliographystyle{ws-procs9x6}

\end{document}